\documentclass[journal=jacsat,manuscript=article]{achemso}

\usepackage[version=3]{mhchem} 

\usepackage{graphicx}
\usepackage{dcolumn}
\usepackage{bm}
\usepackage{comment}
\usepackage{color}
\usepackage{mathtools}
\usepackage{physics}
\usepackage{url} 
\usepackage{achemso}
\setkeys{acs}{articletitle=false, doi=false, keywords=true}

\usepackage{lineno}

\author{Yuki Orimo}
\email{ykormhk@atto.t.u-tokyo.ac.jp}
    \affiliation{Department of Nuclear Engineering and Management, Graduate School of Engineering, The University of Tokyo, 7-3-1 Hongo, Bunkyo-ku, Tokyo 113-8656, Japan}
\author{Takeshi Sato}
    \affiliation{Department of Nuclear Engineering and Management, Graduate School of Engineering, The University of Tokyo, 7-3-1 Hongo, Bunkyo-ku, Tokyo 113-8656, Japan}
    \altaffiliation{Photon Science Center, Graduate School of Engineering, The University of Tokyo, 7-3-1 Hongo, Bunkyo-ku, Tokyo 113-8656, Japan}
  	\alsoaffiliation{Research Institute for Photon Science and Laser Technology, The University of Tokyo, 7-3-1 Hongo, Bunkyo-ku, Tokyo, 113-0033 Japan}
\author{Kenichi L. Ishikawa}
    \affiliation{Department of Nuclear Engineering and Management, Graduate School of Engineering, The University of Tokyo, 7-3-1 Hongo, Bunkyo-ku, Tokyo 113-8656, Japan}
    \altaffiliation{Photon Science Center, Graduate School of Engineering, The University of Tokyo, 7-3-1 Hongo, Bunkyo-ku, Tokyo 113-8656, Japan}
  	\alsoaffiliation{Research Institute for Photon Science and Laser Technology, The University of Tokyo, 7-3-1 Hongo, Bunkyo-ku, Tokyo, 113-0033 Japan}

\title{Efficient simulation of multielectron dynamics in molecules under intense laser pulses: 
Implementation of the multiconfiguration time-dependent Hartree-Fock method based on the adaptive finite element method}

\keywords{\textit{Ab initio} simulation, multielectron dynamics in molecules, intense laser field, TD-MCSCF method}

\begin{document}

\begin{abstract}
We present an implementation of the multiconfiguration time-dependent Hartree-Fock method based on the adaptive finite element method for molecules under intense laser pulses. For efficient simulations, orbital functions are propagated by a stable propagator using the short iterative Arnoldi scheme and our implementation is parallelized for distributed memory computing. This is demonstrated by simulating high-harmonic generation from a water molecule and achieves a simulation of multielectron dynamics with overwhelmingly less computational time, compared to our previous work.
\end{abstract}

\section{\label{sec:intro} Introduction}

Multielectron dynamics studied in strong-field physics and attosecond science is a complicated phenomenon, which includes non-perturbative and nonlinear effects, and multiple states or paths excited by ultrashort pulses~\cite{Brabec_2000,Chang_2011,Calegari_2014}. 
%
\textit{Ab initio} simulations have important roles to understand and predict these physics. Although solving the time-dependent Schrödinger equation (TDSE) gives an exact description of the dynamics in the non-relativistic regime, it is almost impossible to directly solve TDSE for many-body systems due to the exponential growth of the computational cost. The time-dependent multiconfiguration self-consistent field methods (TD-MCSCF) have been developed to overcome this problem~\cite{Ishikawa_2015,Zanghellini_2003,Kato_2004,Caillat_2005,Nguyen-Dang_2007,Miyagi_2013,Sato_2013,Miyagi_2014,Haxton_2015,Sato_2015}. 
In the methods, the total wave function is expressed by the configuration interaction (CI) expansion with time-dependent orbital functions, whose flexibility effectively reduces the required number of configurations. 
The multiconifguration time-dependent Hartree-Fock (MCTDHF) method~\cite{Zanghellini_2003,Kato_2004,Caillat_2005} is the most general approach for fermionic systems. It considers all the possible configurations for a given number of orbital functions.
As further developed methods, the time-dependent complete-active-space self-consistent field method~\cite{Sato_2013}, the time-dependent restricted-active-space self-consistent field method~\cite{Miyagi_2013} and the time-dependent occupation-restricted multiple active-space method~\cite{Sato_2015} have also been proposed. 
They can significantly reduce the number of configurations by classifying orbital functions and making restrictions on electronic excitation.
Today, we can accurately simulate atoms containing several tens of electrons under intense/ultrashort laser pulses thanks to an efficient description of wave functions by TD-MCSCF methods~\cite{Imam_2019}.

However, it is still difficult to handle molecular systems since simple and efficient discretization of the three-dimensional space such as the polar coordinate for atomic systems is not allowed without relying on the symmetries of the systems.
One of the elaborated discretizations to simulate molecules without prohibitive computational cost is using multiresolution grids. The concept of the method is to discretize only a region near nuclei with fine grids and the other regions with grids coarse yet sufficiently fine to describe ionizing wave packets.
We have previously implemented the MCTDHF method based on a multiresolution Cartesian grid and successfully computed high-harmonic generation from a water molecule~\cite{Sawada_2016}.

\begin{figure}[tb]
\begin{center}
\includegraphics[scale=0.5]{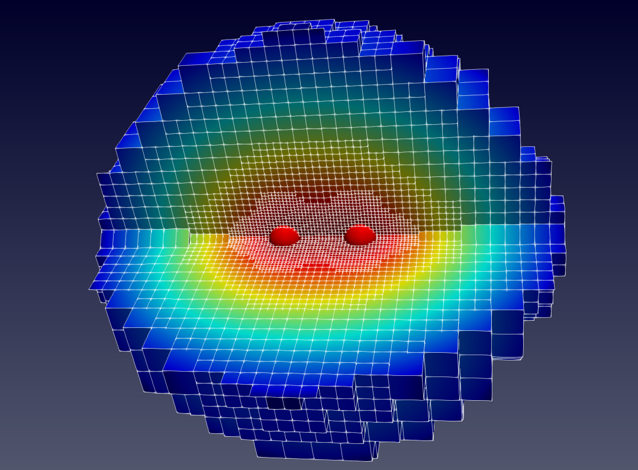}
\caption{A part of an adaptive finite element mesh for a hydrogen molecule. The red spheres show positions of the nuclei and cell colors are electron density.}
\label{h2_grid}
\end{center}
\end{figure}

In this study, we further extend our previous work to implement the MCTDHF method with a finite element method on an adaptively generated multiresolution mesh (adaptive finite element method). As well as our previous implementation, only the center parts of the mesh are refined for sharp changes in wave functions and it gradually becomes coarse in the outer region such as Fig.~\ref{h2_grid}. We can also easily control the order of accuracy since finite element basis functions are used in each cell.
Furthermore, we introduce a highly stable propagator based on the short iterative Lanczos/Arnoldi propagator~\cite{Park_1986} to address instability arising from high spatial resolution.
Our simulation code is parallelized for distributed memory environments, and consequently, achieved over a hundred times faster simulations.

This paper is organized as follows.
In section II, our problem setting is defined and the MCTDHF method is briefly reviewed. In section III, we describe our implementation of spatial discretization using the adaptive finite element method, the time evolution of wave functions with the short iterative Arnoldi propagator, and parallelization. In section IV, we show a numerical result of high-harmonic generation from a water molecule. Conclusions are given in section V.
Hereafter, we use atomic units unless otherwise indicated.

\section{\label{sec:theory} Molecular system and the MCTDHF method}

The Hamiltonian of electrons in a molecule under a laser field can be described as follows.
\begin{linenomath}
\begin{align}
H &= \sum_{i} H_1(\boldsymbol{r_i}) + \frac{1}{2}\sum_{i\neq j} H_2(\boldsymbol{r_i}, \boldsymbol{r_j}) \\
H_1(\boldsymbol{r_i}) &=  -\frac{1}{2}\Delta_i - \sum_a \frac{Z_a}{|\boldsymbol{r}_i - \boldsymbol{r}_a|} - i \boldsymbol{A}(t)\cdot\nabla_i  \\
H_2(\boldsymbol{r_i}, \boldsymbol{r_j})& =  \frac{1}{|\boldsymbol{r}_i - \boldsymbol{r}_j|}
\end{align}
\end{linenomath}
where $\boldsymbol{r}_i$ and $\boldsymbol{r}_a$ are the positions of the $i$th electron and the $a$th nucleus and $Z_a$ is the charge of the $a$th nucleus. 
$\boldsymbol{A}(t) = -\int_\infty^t \boldsymbol{E}(t')dt'$ denotes the vector potential of a laser field applied to the simulated systems, where $\boldsymbol{E}(t)$ is the electric field of it.

Electronic wave functions are modeled by the multiconfiguration time-dependent Hartree-Fock (MCTDHF) method \cite{Zanghellini_2003,Kato_2004,Caillat_2005}. 
Here, we just briefly reviews the method and show the equation of motions (EOMs). The detailed descriptions and derivation of EOMs can be found in the reference~\cite{Sato_2013}.

The MCTDHF method expresses a multielectron wave function $\ket{\Psi}$ with a super position of all the possible Slater determinants composed of a given time-dependent spatial orbital set $\{\phi_p\}$.
\begin{linenomath}
\begin{equation}
    \ket{\Psi} = \sum_I C_I(t) \ket{I}
\end{equation}
\end{linenomath}
$C_I(t)$ is a configuration interaction (CI) coefficient and $\ket{I}$ is an electronic configuration (Slater determinant) composed of orbitals. The equation of motion to variationally evolve the MCTDHF wave function can be derived from the time-dependent variational principle~\cite{Frenkel_1934}. The time-dependent variational principle requires that the action integral $S[\Psi]$, 
\begin{linenomath}
\begin{align}
S[\Psi] &= \int_{t_0}^{t_1} dt \bra{\Psi} \hat{H} - i \frac{\partial}{\partial t} \ket{\Psi},
\end{align}
is stationary to an arbitrary infinitesimal wave function variation $\delta \Psi$,
\end{linenomath}
\begin{linenomath}
\begin{gather}
\frac{\delta S}{\delta \Psi} =  0 \label{stationary_condition}.
\end{gather}
\end{linenomath}
As a solution of the stationary condition (Eq.~\eqref{stationary_condition}), the equations of motion (EOMs) for CI coefficients and orbitals are given as follows.
\begin{linenomath}
\begin{gather}
    i\dot{C}_I = \sum_J \mel*{I}{\hat{H} - i\hat{X}}{J} C_J \label{ci_eom}\\
    \begin{multlined}
    i\ket*{\dot{\phi}_p} = \hat{Q} \left[\hat{H}_1 \ket{\phi_p} +  \sum_{oqrs} (D^{-1})^o_p P^{qs}_{or}  \hat{W}^r_s \ket{\phi_q} \right]+ i \sum_q \ket{\phi_q} X^q_p \label{orb_eom}
    \end{multlined}
\end{gather}
\end{linenomath}
$\hat{X}$ is an arbitrary anti-Hermitian operator, which can be determined as
\begin{linenomath}
\begin{equation}
    \hat{X} = \sum_{pq} X^p_q  \sum_{\sigma}\hat{a}^{\dagger}_{q\sigma} \hat{a}_{p\sigma},
\end{equation}
\end{linenomath}
where $a_{p\sigma} (a^\dagger_{p\sigma})$ is the annihilation (creation) operator for a spatial orbital $\phi_p$ with $\sigma$ spin (up-spin or down-spin), $X^p_q$ is an arbitrary anti-Hermitian matrix. In this work, we set $X^p_q$ to be zero.
$\hat{Q}$ is a projection operator onto the orthogonal complement of occupied orbitals,
\begin{linenomath}
\begin{equation}
    \hat{Q} = 1 - \sum_q \ketbra{\phi_q}.
\end{equation}
\end{linenomath}
$D$ and $P$ are one-body and two-body reduced density matrices, whose matrix elements are defined as
\begin{linenomath}
\begin{gather}
    D^p_q = \sum_\sigma \mel*{\Psi}{ \hat{a}^{\dagger}_{q\sigma} \hat{a}_{p\sigma}}{\Psi}\\
    P^{pq}_{sr} = \sum_{\sigma\tau}  \mel*{\Psi}{\hat{a}^{\dagger}_{s\sigma} \hat{a}^{\dagger}_{r\tau} \hat{a}_{q\tau}\hat{a}_{p\sigma}}{\Psi}.
\end{gather}
\end{linenomath}
$\hat{W}^r_s$ is the inter-electronic mean-field potential given by
\begin{linenomath}
\begin{equation}
    W^r_s(\boldsymbol{r}) = \int d\boldsymbol{r}' \frac{\phi_r^*(\boldsymbol{r}')\phi_s(\boldsymbol{r}')}{|\boldsymbol{r} - \boldsymbol{r}'|}. \label{Eq:mean-field}
\end{equation}
\end{linenomath}

\section{Implementation}
This section shows the implementation of our simulation code developed in this work to solve Eqs.~\eqref{ci_eom} and \eqref{orb_eom} as an initial value problem. Simulations of molecular systems require efficient spatial discretization so that we can simulate electronic dynamics keeping accuracy with realistic computational cost. We employ the adaptive finite element method~\cite{Bangerth_2003,Bangerth_2007} for the efficient discretization of orbitals based on an open-source finite element library deal.II~\cite{Bangerth_2011,dealii2019design}. As described below, while the adaptive finite element method realizes locally high spatial resolution, time evolution could be unstable due to it. To stably propagate wave functions for a long period, We employ the short iterative Arnoldi propagator. Although the short iterative Lanczos propagator is often used in many applications~\cite{Park_1986,Braun_1996,Beck_2000,Feist_2008}, since the system matrix is not Hermitian, the Arnodi algorithm is used instead of the Lanczos algorithm in this application.
Applying this scheme to all orbitals at once, we have enabled more stable time evolution.
These numerical computation schemes are described in the rest of this section.

\subsection{Adaptive finite element method}
The adaptive finite element (AFEM) used in this work is an approach to improve the accuracy of simulations requiring locally high resolution by using a multiresolution mesh generated by local mesh refinement. A finite element mesh is generated by first discretizing the whole simulation box with coarse uniform cubic cells, and then dividing these cells into half the size in regions requiring higher resolution. We can generate an adaptive multiresolution mesh by repeating the second process. Once the multiresolution mesh and cells are generated, most of the rest of the processes fall into the usual finite element method. 

The mesh sizes are determined to make an error in each cell, which is given by Kelly's error indicator ~\cite{Kelly_1983} to estimate the error in each cell from the jump of the gradient of a target function, less than a threshold.
This work adopts the Coulomb potential of the nuclei in a molecule as the target function for the error estimation. We also limit the minimum and maximum mesh sizes to avoid generating extremely small and large cells.

The basis functions located in each cell are direct products of the one-dimensional Lagrange polynomials passing through the Gauss-Lobatto quadrature points in each cell. The quadrature points in each cell are also constructed as the direct product of one-dimensional Gauss-Lobatto quadrature points. This basis can be considered to be the three-dimensional version of the finite element discrete variable representation (FEDVR) basis \cite{Rescigno_2000,McCurdy_2004}.

Let us define $f_{I,i}(\boldsymbol{r})$ as the $i$ th basis function in the $I$ th cell, and $L_{I,j_x} (x)$,
$L_{I,j_y} (y)$ and $L_{I,j_z} (z)$ the $(j_x, j_y, j_z)$ th Lagrange polynomials in each dimension in the $I$ th cell.
Then, the function $f_{I,i}(\boldsymbol{r})$ is given by
\begin{linenomath}
\begin{equation}
    f_{I,i}(\boldsymbol{r}) = L_{I,j_x} (x) L_{I,j_y} (y) L_{I,j_z} (z).
\end{equation}
\end{linenomath}
These functions are defined only in the $I$ th cell and have zero values in other region than that.


The finite element basis set $\{b_k(\boldsymbol{r})\}$ is constructed by the basis functions $f_{I,i}(\boldsymbol{r})$ which have zero-value on the boundary of each cell and bridged functions that combine two bases with nonzero values at the quadrature point shared by two cells on the boundary of adjacent cells. 
The bridged functions are required to ensure the continuity of discretized functions.
The mesh generation and construction of the basis are carried out by using deal.II functions.

An arbitrary function $h(\boldsymbol{r})$ is discretized with this finite element basis as follows.
\begin{linenomath}
\begin{gather}
    h(\boldsymbol{r}) = \sum_k c_{k} b_k(\boldsymbol{r})\\
    c_{k} = \sum_l (\tilde{M}^{-1})_{k,l}\int d\boldsymbol{r} b_l(\boldsymbol{r}) h(\boldsymbol{r})
\end{gather}
\end{linenomath}
The matrix $\tilde{M}$ is the overlap matrix of the basis set $\{b_k(\boldsymbol{r})\}$, called the mass matrix in the finite element method, defined as
\begin{linenomath}
\begin{equation}
    \tilde{M}_{k,l} = \int d\boldsymbol{r} b_k(\boldsymbol{r}) b_l(\boldsymbol{r}).
\end{equation}
\end{linenomath}
All the spatial integrals are approximated with Gauss-Lobatto quadrature as follows.
\begin{linenomath}
\begin{equation}
    \int d\boldsymbol{r} h(x,y,z)\simeq \sum_I \sum_{j_x, j_y, j_z} w^x_{I,j_x} w^y_{I, j_y} w^z_{I,j_z} h(x_{I,j_x}, y_{I,j_y}, z_{I,j_z}),
\end{equation}
\end{linenomath}
where $w^d_{I,j_d}~(d = x, y, z)$ and $(x_{I,j_x}, y_{I,j_y}, z_{I,j_z} )$ are the quadrature weights and points of the $I$ th cell.

Based on this discretization scheme, the equation of motion (Eq.~\eqref{orb_eom}) is converted into a matrix-vector equation,
\begin{linenomath}
\begin{equation}
    i \tilde{M} \dot{\boldsymbol{c}}_p = (1-\tilde{M}  \sum_q  \boldsymbol{c}_q  \boldsymbol{c}^\dagger_q) \left[ \tilde{H}_1 \boldsymbol{c}_p + \tilde{M}\sum_{oqrs} (D^{-1})^o_p P^{qs}_{or} \boldsymbol{W}^r_s \circ \boldsymbol{c}_q \right]
    + i \tilde{M}  \sum_q \boldsymbol{c}_q X^q_p \label{Eq:discretized_eom}
\end{equation}
\end{linenomath}
where $\boldsymbol{c}_p$ denotes a coefficient vector of orbital $\phi_p(\boldsymbol{r})$ given by,
\begin{linenomath}
\begin{gather}
    (\boldsymbol{c}_p)_k = \int d\boldsymbol{r} b_k(\boldsymbol{r}) \phi_p(\boldsymbol{r})
\end{gather}
\end{linenomath}
and the matrices $\tilde{H}_1$ is defined as the matrix form of the operator $\hat{H}_1$,
\begin{linenomath}
\begin{gather}
    (\tilde{H}_1)_{k,l} = \int d\boldsymbol{r} b_k(\boldsymbol{r}) H_1(\boldsymbol{r}) b_l(\boldsymbol{r}).
\end{gather}
\end{linenomath}
$\boldsymbol{W}^r_s$ is a coefficient vector of the mean-field potential $W^r_s(\boldsymbol{r})$ and the element-wise product is denoted by ``$\circ$".

We compute the mean-field potential by solving the following Poisson's equation, instead of directly calculating Eq.~\eqref{Eq:mean-field},
\begin{linenomath}
\begin{gather}
    \Delta W^r_s(\boldsymbol{r}) = -4\pi \phi_r^*(\boldsymbol{r}) \phi_s(\boldsymbol{r})
\end{gather}
\end{linenomath}
with a boundary condition
\begin{linenomath}
\begin{equation}
    W^r_s(\boldsymbol{r}) \Big|_{\boldsymbol{r} \in \Omega} = \int d\boldsymbol{r}' \frac{\phi_r^*(\boldsymbol{r}')\phi_s(\boldsymbol{r}')}{|\boldsymbol{r} - \boldsymbol{r}'|},
\end{equation}
\end{linenomath}
where $\Omega$ denotes the boundary of a simulation box.
We solve this equation by the conjugate gradient method with algebraic multigrid preconditioning implemented in an open-source parallel linear algebra library Trilinos~\cite{trilinos-website} interfaced on deal.II.


\subsection{Short iterative Arnoldi propagator}
The short iterative Lanczos/Arnoldi propagator is a time evolution method, which approximates a Hamiltonian in a Krylov subspace by the Lanczos/Arnoldi algorithm and iterates short-time propagation of wave functions in the subspace~\cite{Park_1986}.
This approach conserves the norm of a wave function when a Hamiltonian is Hermitian and enables unconditionally stable time evolution. It is also possible to use an adaptive time step or a variable Krylov subspace dimension based on the error estimation \cite{Park_1986}
However, we cannot straightforwardly apply it to the equation of motion of orbitals, since it is only applicable to linear equations.

Although some applications of the MCSCF methods, where the EOM of orbitals is nonlinear, use exponential integrators~\cite{Hochbruck_2010,Auziger_2020,Gomez_2018} to enjoy the stability of the short iterative Lanczos/Arnoldi propagator even only for linear parts of the EOM, in our application, we found that the explicit time propagation of the nonlinear parts causes numerical instability probably due to the quite fine mesh of AFEM.
To avoid this problem, in this work, we propose an approach to apply the short iterative Lanczos/Arnoldi propagator by approximately regarding the whole of the EOMs for all orbitals as one linear system.

The equations of motion for all orbitals (Eq.~\eqref{orb_eom}) can be packed into a matrix-vector form, whose elements are operators and ket-vectors.
\begin{linenomath}
\begin{gather}
    i \frac{\partial}{\partial t} 
    \boldsymbol{\phi}
    = 
    \hat{G}
    \boldsymbol{\phi} \label{packed_orb_eom}
    \\
    \boldsymbol{\phi} = 
    \begin{bmatrix}
    \ket{\phi_1} \\ \ket{\phi_2} \\ \vdots \\ \ket{\phi_n}
    \end{bmatrix}, \quad
    G = 
    \begin{bmatrix}
        \hat{G}_{11} & \hat{G}_{12} & \cdots & \hat{G}_{1N}\\
        \hat{G}_{21} & \hat{G}_{22} & \cdots & \hat{G}_{2N}\\
        \vdots &                    & \ddots & \vdots\\
        \hat{G}_{N1} & \hat{G}_{N2} & \cdots & \hat{G}_{NN}\\
    \end{bmatrix}
\end{gather}
\end{linenomath}
The matrix element $\hat{G}_{ij}$ is an operator defined as 
\begin{linenomath}
\begin{equation}
    \hat{G}_{ij} = \delta^i_j \hat{H}_1 + \sum_{osr} (D^{-1})^o_i P^{js}_{or} \hat{W}^r_s 
    -\bra{\phi_j} \left[ \hat{H}_1 \ket{\phi_i} + \sum_{oqrs} (D^{-1})^o_i P^{js}_{or} \hat{W}^r_s \ket{\phi_q} \right] + i X^j_i.
\end{equation}
\end{linenomath}
The equation~\eqref{packed_orb_eom} is approximately linear if we can assume that orbitals in the operators are invariable within a short time $\Delta t$, and then time evolution of orbitals can be described as 
\begin{linenomath}
\begin{equation}
    \boldsymbol{\phi}(t+\Delta t)
    = \exp(-i\hat{G}\Delta t) \boldsymbol{\phi}(t).
\end{equation}
\end{linenomath}
We achieve this time evolution by the short iterative Arnoldi scheme.
Although this scheme has first-order accuracy since the time-dependency of the operator $\hat{G}$ in a time step $\Delta t$ is not considered, it enables highly stable propagation including nonlinear parts and fits our implementation based on the AFEM using a fine mesh.
The Krylov subspace dimension of the Arnoldi algorithm is determined so that errors estimated by the method found in the references~\cite{Park_1986,Beck_2000} are less than a threshold, which is set to be $10^{-10}$ in this work. We also adjust a time-step size, which is fixed during our simulations, to make the dimension 10-15 at a maximum. 


\subsection{Parallelization}
The spatial discretization and time evolution discussed above are devised to efficiently simulate multielectron dynamics in a laser field. Nevertheless, computational costs for the three-space to describe laser-induced ionization are huge , and distributed memory parallel computing is essential.
%
The total number of degrees of freedom (DOF) $N_\text{DOF}$ in our simulation can simply be written as $N_\text{DOF} = N_\text{orbital}\times N_\text{space} + N_\text{CI}$, where $N_\text{orbital}$, $N_\text{space}$ and $N_\text{CI}$ are the numbers of orbitals, DOF associated with spatial discretization and CI coefficients, respectively. $N_\text{orbital}$ is typically from several to several tens, and  $N_\text{space}$ usually increases up to several millions. $N_\text{CI}$ significantly changes depending on a problem since it exponentially increases to the numbers of electrons and orbitals.
Our strategy to make efficient use of many processors in many situations is parallelizing orbital functions with respect to both the number of orbitals and the number of degrees of freedom in the AFEM.
%

\begin{figure}[tb]
\begin{center}
\includegraphics[scale=0.65]{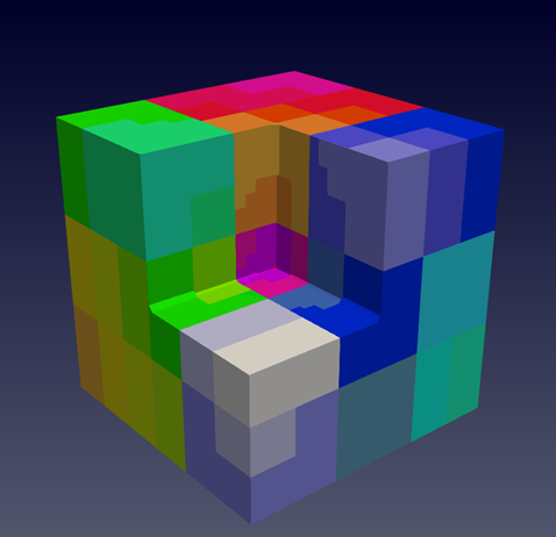}
\caption{An example of a divided simulation box. The color-coded cells correspond the distribution to MPI processes.}
\label{distributed_space}
\end{center}
\end{figure}

We divide the orbital function set $\{\ket{\phi_p}\}$ by $K$ and create $K$ MPI groups to deal with them. Each MPI group has $L$ independent processes that are used to distribute a simulation box by using deal.II functions.
Distribution of a simulation box and DOFs accompanying it is carried out by p4est~\cite{Burstedde_2011,Bangerth_2011}, an open-source library to distribute octree structures across multiple processors, interfaced to deal.II. This addresses load balancing and optimal distribution of the simulation box to reduce MPI communications among the processors (Fig.~\ref{distributed_space}).

\section{\label{sec:applications} Applications}
\begin{figure}[tb]
\begin{center}
\includegraphics[scale=0.85]{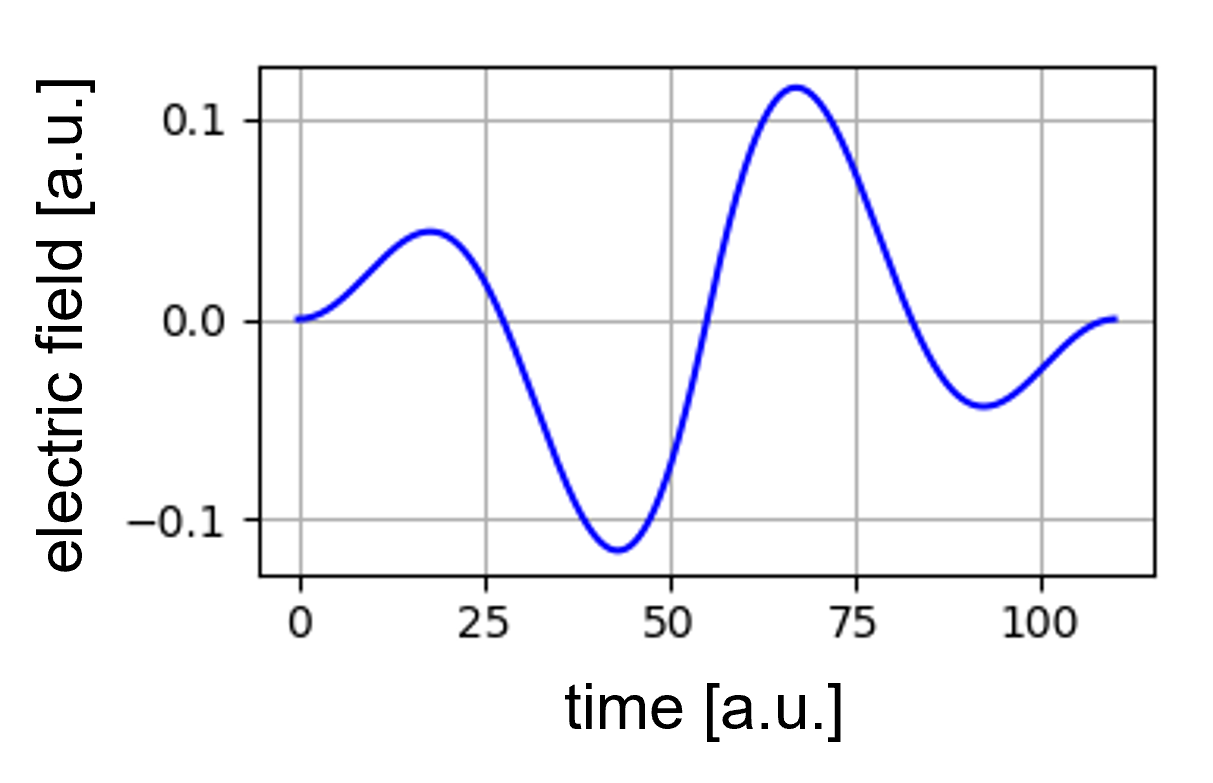}
\caption{The electric field of the laser pulse used in this simulation.}
\label{sim1_laser}
\end{center}
\end{figure}

We simulate high harmonic generation from a water molecule to demonstrate the efficiency of our implementation by comparing our previous work~\cite{Sawada_2016}.
For atomic positions of a water molecule, two hydrogen atoms of a water molecule are located at $(\pm1.42994, 1.10718, 0)$ and an oxygen atom is located at the origin.
The laser pulse used in this simulation has a wavelength of $2\pi c/\omega = 400 \mathrm{nm}$ ($c$ is the speed of light in vacuum) and a peak intensity of $I_0 = 8 \times 10^{14}~\mathrm{W/cm^2}$, and is linearly polarized along with $x$-axis.
The pulse duration is 2 optical cycles with a triangular envelope.
The shape of the electric field of the laser pulse is defined as, (see also Fig.~\ref{sim1_laser}),
\begin{linenomath}
\begin{align}
    E(t) &= E_0 f_\mathrm{env}(t) \sin(\omega t)\\
    f_\mathrm{env}(t) &= 
    \begin{cases}
    \displaystyle \frac{\omega t}{2\pi}  & (0 \leq \omega t \leq 2\pi)\\[12pt]
    \displaystyle \frac{4\pi - \omega t}{2\pi}  & (2\pi \leq \omega t \leq 4\pi)\\
    \end{cases},
\end{align}
\end{linenomath}
where $E_0$ is the peak electric field derived from the peak intensity.
The time-step size for real-time evolution is 0.01~a.u..

The simulation box is a cuboid defined within a region $[-70, 70]\times[-30, 30]\times[-30, 30]$.
We apply the exterior complex scaling (ECS) as an absorbing boundary in the outside of a region $[-35,35]\times[-10, 10]\times[-10, 10]$. The details of the ECS can be found in the references~\cite{McCurdy_1991,Scrinzi_2010,Orimo_2018}. 

\begin{figure}[tb]
\begin{center}
\includegraphics[scale=0.6]{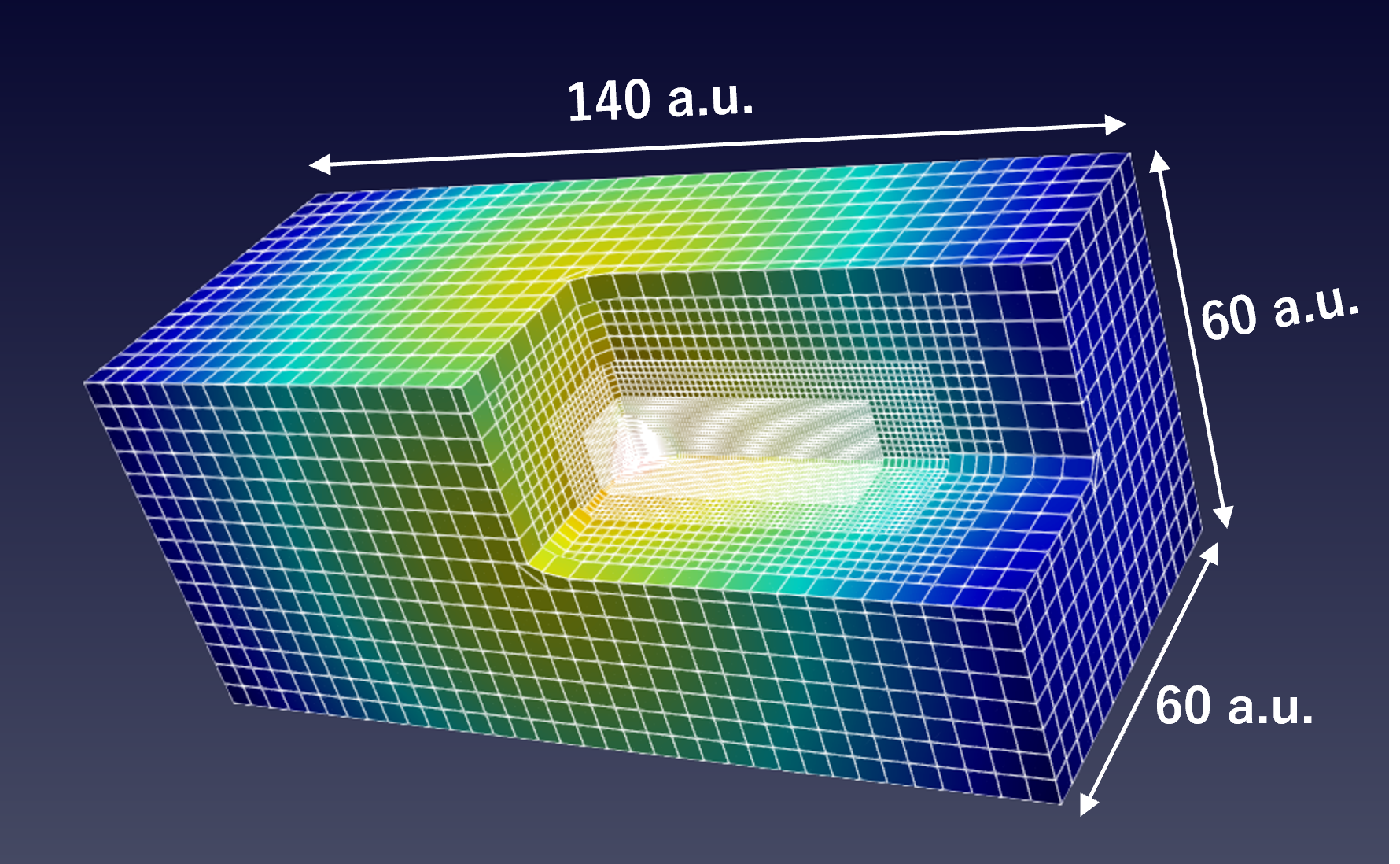}
\caption{Adaptively generated finite element mesh for a water molecule. The largest element is a cube of edge length 4.0~a.u. used to discretize the outer region, and  the smallest one is a cube of edge length 0.125~a.u. used only in the vicinity of nuclei.}
\label{h2o_grid}
\end{center}
\end{figure}

The finite element mesh is generated to satisfy that the error in each cell is less than 0.005, which has 6 different sizes between $0.125~\mathrm{a.u.}$ and $4.0~\mathrm{a.u.}$.
At the most distant region from the molecule, the largest elements, which are cubes with 4.0~a.u long sides, are used to describe sufficiently absorbed orbital functions and the smallest elements, whose edge length is 0.125~a.u., are used in the vicinity of the molecule. Figure~\ref{h2o_grid} displays the finite element mesh used in this simulation.
The finite element basis is constructed from first-order Lagrange polynomials, and thus there are 8 quadrature points in a finite element cell.
While it is possible to dynamically adapt a mesh to time-dependent orbital functions, we avoid such approaches due to additional computational costs. This would be helpful to gain computational efficiency if our problem was a larger system.

\begin{figure}[tb]
\vspace{10pt}
\begin{center}
\includegraphics[scale=0.55]{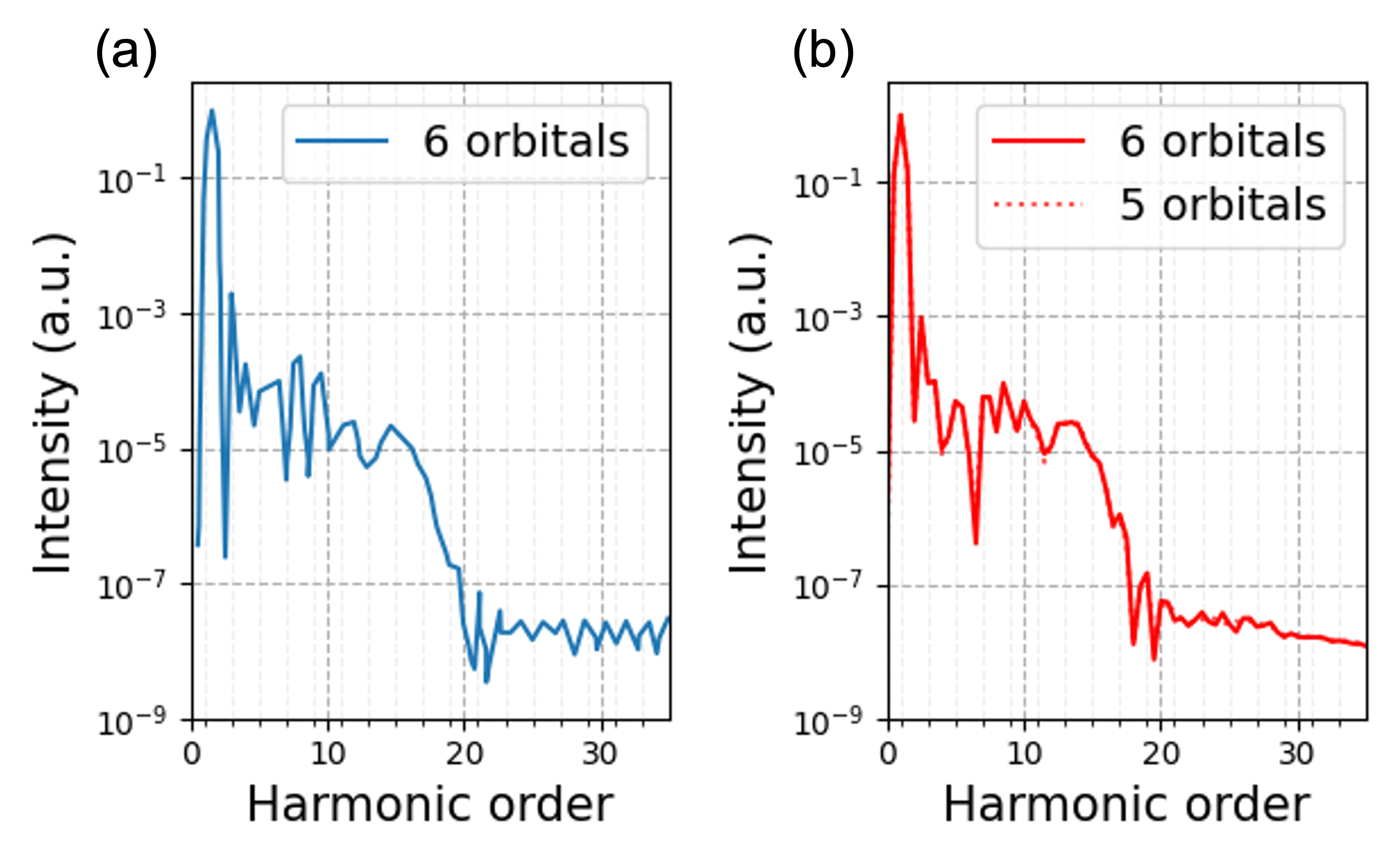}
\caption{High harmonic spectra of a water molecule exposed to a laser pulse with a wavelength of $400 \mathrm{nm}$ and a peak intensity of $8 \times 10^{14}~\mathrm{W/cm^2}$. (a) The spectrum taken from Ref.~\cite{Sawada_2016}. The data is normalized for the maximum to be unity. (b) The spectra computed by the present work.}
\label{h2o_hhg_compare}
\end{center}
\end{figure}

For the beginning of the simulation, we computed a ground state by imaginary-time evolution, whose electronic energy was -76.905~a.u.
In figure.~\ref{h2o_hhg_compare}, we compare our simulation result with the previously calculated one.
These spectra do not perfectly agree with each other since it is extremely difficult to achieve perfect convergence for spatial resolutions in 3D systems, Nevertheless, overall spectral shapes are quite similar.
As well as the previous calculations, the simulations with 5 orbitals and 6 orbitals give almost the same spectra.
The simulation using 6 orbitals of present work took 6.5 hours with 240 cores (6 nodes, 2 Intel Xeon Gold 2.40GHz processors with 20 cores in a node). 
Remarkably, it is about 100 times faster than the previous work which took 28 days to finish the simulation.
One of our achievements is successful distributed parallel computing using the 20 times larger resource.
In addition to this, at least 5 times acceleration was gained by factors except for parallelization.
The development of a highly stable propagator mainly contributes to this speed-up, which enables time evolution with 4 times as large a time-step size as the previous one.

\section{Conclusion}
We have implemented the MCTDHF method based on the adaptive finite element method to simulate multielectron dynamics in molecules under laser fields. 
A further sophisticated discretization is realized by using the multiresolution grid used in our previous implementation in the frame of the finite element method. Thanks to the finite element method, we can automatically generate an adaptive mesh using Kelly's error indicator and easily control the order of accuracy by changing the polynomial order of basis functions.
While locally refined meshes enable efficient and accurate simulations, they possibly make time evolution unstable.
We developed a more stable propagator based on the short iterative Arnoldi scheme than exponential integrators. 
This propagator evolves all orbital functions together as a vector by using the short iterative Arnoldi scheme.
In addition, our simulation code is parallelized for distributed memory computing, which handles both the orbital set and spatial degrees of freedom in parallel.

We have applied the present implementation to a simulation of high-harmonic generation from a water molecule in an intense visible laser pulse to compare with our previous work~\cite{Sawada_2016}, and obtained the spectra showing a good agreement with overwhelmingly less computational time.
Parallelization has made the greatest contribution to this reduction in computation time, and in this study, we were able to successfully use 20 times larger computational resources than in the past. It is also important to note that we were able to use a 4 times larger time-step size thanks to the stable propagator.

This study prepared the adaptive mesh based on the discretization error of the Coulomb potential of the nuclei, therefore the mesh is fixed during simulations, but it is possible to dynamically adapt the mesh to a wave function or nuclear positions at each time step.
We consider that it brings efficiency when a larger simulation box is needed or when the nuclei can move.
In future works, we will present \textit{ab initio} simulations of more complicated molecular systems and simulations considering nuclear dynamics in a combination of this development and more advanced theories such as the TD-ORMAS method~\cite{Sato_2015} and the time-dependent coupled cluster theory~\cite{Sato_2018}.

\section*{Data availability}
The data and source code used in this study are available upon reasonable request.

\section*{Competing interests}
The authors declare there are no competing interests.

\section*{Funding information}
This research was supported in part by a Grant-in-Aid for Scientific Research (Grants No. JP19H00869, No. JP21K18903, and No. JP22H05025) and a Grant-in-Aid for Early-Career Scientists (Grant No. JP22K14616) from the Ministry of Education, Culture, Sports, Science and Technology (MEXT) of Japan. This research was also partially supported by JST CREST (Grant No. JPMJCR15N1) and by MEXT Quantum Leap Flagship Program (MEXT Q-LEAP) Grant Number JPMXS0118067246.

\bibliographystyle{achemso}
\bibliography{all10.bib,all13.bib}

\end{document}